\documentstyle[preprint, prb,aps,epsf]{revtex}
\begin{document}

\newcommand{\threeby}{3$\times$3}
\newcommand{\etaMnN}{Mn$_{3}$N$_{2}$}
\newcommand{\MnxNy}{Mn$_{x}$N$_{y}$}
\newcommand{\Jmn}{{\it J}$_{Mn}$}
\newcommand{\Jn}{{\it J}$_{N}$}
\newcommand{\etaparallel}{$\eta_{\parallel}$}
\newcommand{\Fn}{{\it f}$_{N}$}
\newcommand{\Fmn}{{\it f}$_{Mn}$}
\newcommand{\Fv}{{\it f}$_{V}$}

\title{Energy Dependent Contrast in Atomic-Scale Spin-Polarized
Scanning Tunneling Microscopy of \etaMnN\ (010): Experiment and First-Principles Theory}
\author{Rong Yang\footnote{Corresponding author1, yang@helios.phy.ohiou.edu}, Haiqiang Yang and
Arthur R. Smith\footnote{Corresponding author2, smitha2@ohio.edu}\\ \small{Nanoscale and Quantum Phenomena
Institute, Department of Physics and Astronomy, Ohio University, Athens, OH 45701}\\ }
\author{Alexey Dick and J{\"o}rg Neugebauer\\ \small{Max-Planck-Institut f{\"u}r Eisenforschung GmbH, Max-Planck Str. 1,
40237 D{\"u}sseldorf, Germany\\ Theoretische Physik, Universit{\"a}t Paderborn, Warburger Str. 100, 33098
Paderborn, Germany}\\}

\maketitle

\begin{abstract}
The technique of spin-polarized scanning tunneling microscopy is investigated for its use in determining fine
details of surface magnetic structure down to the atomic scale. As a model sample, the row-wise anti-ferromagnetic
\etaMnN (010) surface is studied. It is shown that the magnetic contrast in atomic-scale images is a strong function
of the bias voltage around the Fermi level. Inversion of the magnetic contrast is also demonstrated. The experimental
 SP-STM images and height profiles are compared with simulated SP-STM images and height profiles based on spin-polarized
 density functional theory. The success of different tip models in reproducing the non-magnetic and magnetic STM data is
 explored.
\end{abstract}

{\bf PACS:} 68.37.Ef, 75.70.-i, 75.50.Ee, 75.70.Rf

\newpage

\subsection{Introduction}
Nanoscale magnetism is a topic of increasing interest, having potential applications in advanced data storage and
spin-based electronics. One of the most powerful techniques with the ability to probe magnetic structure at the
nanoscale is spin-polarized scanning tunneling microscopy (SP-STM). This technique offers spin sensitivity combined
with the well-known advantage of STM - namely, spatial resolution down to the atomic-scale. SP-STM has thus far
been utilized to great advantage in imaging both ferromagnetic and antiferromagnetic surface
structures\cite{Wiesendanger1,Wiesendanger2,Kleiber,Heinze,YangPRL,Okuno,Bode1,Yamada,Kubetzka1,SmithSS,Kubetzka2}.

The principle of SP-STM is based on the variation of the tunneling current with the angle between the spin of the
tip and that of the sample - larger current for parallel orientation and smaller current for antiparallel
orientation. This effect depends on the local density of states for spin-$\uparrow$ and spin-$\downarrow$ channels.
SP-STM has been used by Wiesendanger {\it et al.} to map the energy-dependent spin polarization of nanoscale
ferromagnetic Gd(0001) islands by applying a magnetic field to the sample and measuring dI/dV vs. V in the
tip-sample parallel and antiparallel configurations.\cite{Wiesendanger2} For deriving the polarization in the case
of an antiferromagnetic (aFM) surface, an external field is not required since the parallel and antiparallel
configurations occur periodically as one moves the tip from row to row.

Regarding aFM surfaces, Heinze {\it et al.} reported in 2000 the atomic-scale magnetic contrast of a Mn monolayer
on a W(110) using atomic-scale SP-STM in constant current (CC) mode.\cite{Heinze} Two years later, Yang {\it et
al.} showed the simultaneous magnetic and non-magnetic contrast on \etaMnN\ (010) using also atomic-scale SP-STM in
CC mode.\cite{YangPRL} Very recently, Kubetzka {\it et al.} have applied the same technique to study a single
monolayer of Fe on W(001), resolving a longstanding question regarding the magnetic ground state of this
system.\cite{Kubetzka2} Atomic-scale SP-STM has therefore been established as a powerful technique for resolving
the ultimate limits of magnetic structure on surfaces.

In this article, we explore the effect of bias voltage on the magnetic corrugation on aFM  \etaMnN\ (010). The
\etaMnN (010) surface is a model aFM surface for such studies. The Mn atoms of the highly bulk-like surface have
magnetic moments which alternate in a row-wise aFM arrangement.\cite{SmithSS,YangSS} Bulk measurements show the
corresponding layer-wise aFM structure with magnetic moments of Mn1 and Mn2 which are each close to $\sim$ 3.5
$\mu_B$,\cite{Leineweber,YangAPL} and bulk theoretical calculations are in good agreement.\cite{Walter1} SP-STM
corrugation profiles in general are found to contain both magnetic and non-magnetic information. Here, it is found
that the magnetic STM image depends strongly on the bias voltage, and at certain voltage the magnetic amplitude
goes to zero. The bias voltage-dependence is also found to affect the SP-STM line profile shape. Comparisons of the
experimental SP-STM images and height profiles to simulated SP-STM images and height profiles calculated from
spin-polarized first-principles density functional theory are shown. In particular, comparisons are made between
the non-magnetic and magnetic STM profiles for experiment vs. theory utilizing various tip electronic structures,
including constant DOS tip and Fe-atom tip. We compare the relative importance of the detailed tip electronic
structure on the non-magnetic vs. magnetic STM simulations.

\subsection{Experimental Approach}
The experiments are performed in a combined ultra high vacuum (UHV) growth/analysis system which allows direct UHV
transfer of samples from the growth chamber to the STM chamber. The MBE system includes a solid source effusion
cell for Mn and a RF plasma source for N. After being heated up to 1000 $^\circ$C for 30 minutes with the nitrogen
plasma turned on, the MgO substrate temperature is lowered to 450 $^\circ$C prior to the growth of manganese
nitride \etaMnN (010) thin film. The nitrogen flow rate is about 1.1 sccm (growth chamber pressure is
1.1$\times$10$^{-5}$ Torr) with the RF power set at 500 W. The Mn flux is about 3.5$\times$10$^{14}$/cm$^{2}$s. The
growth condition is monitored using reflection high energy electron diffraction (RHEED)\cite{YangAPL}.

Following growth, the samples are investigated with {\it in situ} SP-STM. For normal STM studies, we use
electrochemically etched tungsten tips which are cleaned in the ultra high vacuum chamber using electron
bombardment. For SP-STM studies, we coat the cleaned tips with Fe at room temperature to a thickness of 5-10
monolayers. Coated tips are magnetized in a small $\sim$40 mT magnetic field directed perpendicular to the tip axis.
SP-STM images are taken in the constant current (CC) mode at room temperature.

\subsection{Theoretical Approach}
In order to understand theoretically the magnetic surface structure of \etaMnN\ (010), we have performed an {\it
ab-initio} study. To describe the SP-STM experiments, we have used the spin-polarized Tersoff-Hamann approach. The
tunneling current can be expressed as \cite{Wortmann}

\begin{equation}{}
I_{t}\sim\int dE\, g_{v}(E)\,n^{t}\left(E-eV_{Bias}\right)\left[
 n^{s}\left(\{\mathbf{R}_{t}\},E\right)+\frac{m^{t}\left(E-eV_{Bias}\right)}{n^{t}\left(E-eV_{Bias}\right)}\,\cdot\,\mathbf{m}^{s}\left(\{\mathbf{R}_{t}\},E\right)\right]
\end{equation}

Here, $g_{v}\left(E\right)=f\left(E-E_{F}\right)-f\left(E-E_{F}-eV_{Bias}\right)$, $f$ is the Fermi function,
$E_{F}$ is Fermi energy of the sample, and $n^{t}\left(E\right)$, $\mathbf{m}^{t}\left(E\right)$,
$n^{s}\left(\{\mathbf{R}_{t}\},E\right)$, $m^{s}\left(\{\mathbf{R}_{t}\},E\right)$ are energy-dependent
non-magnetic and magnetic densities of states (DOS) of the tip and local densities of states (LDOS) of the sample
at the tip position $\mathbf{R}_{t}$ respectively.
$\frac{m^{t}\left(E\right)}{n^{t}\left(E\right)}=P_t\left(E\right)$ is the energy dependent magnetic polarization
of the tip.

The non-magnetic $n^{s}\left(\{\mathbf{R}_{t}\},E\right)$ and magnetic $m^{s}\left(\{\mathbf{R}_{t}\},E\right)$
LDOS of the sample entering Eq. 2 have been calculated employing density functional theory within the spin-LDA
approximation. We have used a plane-wave pseudopotential approach as implemented in SFHIngX \cite{WWW}. Details of
the calculations are described elsewhere \cite{SmithSS}. To describe the surface electronic structure a plane-wave
cutoff energy of 50Ry and a slab consisting of four layers has been used. All calculations have been performed with
a bias independent constant tunneling current I$_{t}(eV_{Bias})$=I$_{const}$, to model the experimental constant
current mode.

Since the exact geometric, electronic, and magnetic properties of the tip used in the experiment are not known (and
can even change during the experiment), we employed tip models having different electronic properties to verify the
effect of the tip electronic structure on the final SP-STM profiles. We have started with the conventional approach
assuming a magnetic tip with an energy-independent DOS, and compared these results with a more realistic Fe-atom
tip. The DOS for Fe tip atoms has been extracted from work by Heinze \emph{et al.}\cite{Heinze1}.

As has been shown previously \cite{SmithSS} for the case of -0.2V bias, a multi-atom tip apex provides a better
agreement between calculated and measured profiles. According to this model, all apex atoms are assumed to have
predominant $s$-type electronic character and to not interfere with each other. Then, each apex atom contributes to
the total tunneling current according to Eq. 1.

In the present study, we have also employed a tip broadening model. The model tip consists of an apex of four
atoms, sitting at the corners of a square parallel to the surface. The lateral apex atom-to-atom distance d$_{t}$
is assumed bias-independent. The sensitivity of the theoretical non-magnetic profiles to the tip geometry
parameters d$_{t}$ and I$_{const}$ has been carefully analyzed for case of constant DOS tip. We have varied the tip
geometry between 1.40\AA\ $\leq$ d$_{t}$ $\leq$ 1.85\AA. In all cases, the average error between measured and
simulated non-magnetic magnitude lies within a range of 10\%-25\% and is found to have no effect on the qualitative
shape of the profiles.

\subsection{Results and Discussion}
\subsubsection{Magnetic Contrast on \etaMnN\ (010)}
Previous work has elucidated the geometrical structure of the (010) surface, which is depicted in Fig. 1(a) in top
view.\cite{YangAPL} The surface is composed of Mn atoms in a face-centered tetragonal structure. There are two
types (configurations) of surface Mn atoms - Mn1, which have 2 in-plane bonds to N; and Mn2, which have 3 in-plane
bonds to N. Since the bulk {\it c}-planes are layer-wise antiferromagnetic and the {\it c}-axis is in the surface
plane, the surface atom magnetic moments alternate in a row-wise aFM arrangement. The atomic row-spacing along
[001] is 2.02 \AA, whereas the row-spacing along [100] is 2.10 \AA. Figure 1(b) illustrates the geometrical
configuration for the SP-STM experiment.

Experimental CC-mode SP-STM images of the \etaMnN\ (010) surface have been obtained, as shown in Fig. 2. This image
is of size 100 \AA$\times$100 \AA\ with tunneling current I$_T$ = 0.3nA and at sample bias voltage V$_S$ = -0.2V.
The image shows the row structure of the \etaMnN\ (010) surface; the spacing of the rows is 6.07 \AA\ which is
equal to $c$/2 or 3 single atomic rows. The row spacing is therefore that between rows of Mn1 atoms.  A number of
defects, probably Mn1 vacancies, are observed on the surface. Although chemically all the Mn1 rows are the same, a
uniform modulation of the row-heights is seen, which is due to the spin-polarized effect.\cite{YangPRL} Since the
STM is operated at constant current, the tip has to withdraw from or come closer to the surface depending on the
magnetization directions of tip and sample. For Cr(001), this led to an alternation of the apparent step
height.\cite{Wiesendanger1} For a row-wise aFM surface, CC mode leads to the periodic modulation of the peak
heights. The period of the magnetic modulation is two Mn1 rows, or c=12.14\AA\ .

\subsubsection{Analysis of the Bias Dependence of the Magnetic Contrast}
Figure 3(a-e) and Fig. 4(a-e) show sets of CC mode SP-STM images acquired from the exact same location on the
surface, namely from the boxed region shown in Fig. 2 (we used the defects on the surface to locate the position).
As with the larger image shown in Fig.2, these STM images do not resolve the individual Mn1 and Mn2 atoms; this is
common for all of our spin-polarized STM images of this surface. It suggests that the magnetic tips are less sharp
than a single atom tip. Note that we have published a complete study of the bias-dependence of the
atomic-resolution non-magnetic images acquired with a sharp (possibly single atom) non-magnetic tip.\cite{YangSS}
In that data set, the Mn1 and Mn2 atoms are clearly resolved. Since there is no Mn1 and Mn2 atom resolution in the
data shown here, the tip broadening model as described in Sec.C has been used for the simulations.

The [100]-averaged height profiles are also shown below each image and on the same scale throughout, where the
height modulation of the rows as well as the overall corrugation can be clearly observed. What we find from these
images is that there are distinct variations of the line profile as a function of energy, and several key points
can be made. First, the overall corrugation magnitude is largest at the smallest bias magnitudes; we note that this
trend is consistent with the atomic-resolution bias-dependent images.\cite{YangSS} Second, the magnetic modulation
reaches a maximum of $\sim$ 0.04 \AA\ at $V_S$ = -0.1V. Third, magnetic modulation is observed clearly at all
energies except at $V_S$ $\sim$ +0.4V, where it becomes very small. In fact and as the fourth key point, at $V_S$
$\sim$ +0.4V, the modulation undergoes a reversal. This magnetic contrast reversal is clear by counting the number
of "high peaks" and "low peaks". For $V_s$ $<$ 0.4 V, there are 2 high peaks (indicated by upward red arrows in
Figs. 3 and 4) and 3 low peaks, whereas for $V_s$ $>$ 0.4 V, there are 3 high peaks and 2 low peaks.

We have separated the magnetic and non-magnetic components from the total SP-STM experimental height profiles. This
is done by adding and subtracting, for a given image, the averaged line profile and the same line profile shifted
by half the magnetic period = {\it c}/2, as explained in more detail elsewhere.\cite{YangPRL} The results are shown
in Fig. 5(a-e) and Fig. 6(a-e) for each image corresponding to Figs. 3(a-e) and 4(a-e). It is clear that at all
energies, the non-magnetic component has a smooth sinusoid-like shape.

On the other hand, the magnetic component has a shape which varies with the energy. While to first-order, the
magnetic profile shape is roughly sinusoidal, at many voltages within the range from -0.8V to +0.2V the magnetic
profile shows a distinctly trapezoid-like shape. This is very clear at, for example, -0.6V and -0.2V in Fig. 5. At
positive voltages greater than 0.2V, this trapezoidal shape is not evident; the profile is more rounded. Of course,
at $V_S$ $\sim$ 0.4V, the magnetic line profile is nearly flat, and the contrast reversal is near that point.

\subsubsection{The Necessity of Energy-Dependent Tip Polarization}
To analyze the SP-STM experiments, we have modeled the bias-dependent profiles using Eq. 1 and employing the
conventional assumption of a constant (with energy) tip density of states, as was done, e.g. in
Refs.\cite{Bode1,Wiesendanger2,Tedrow}, where it was assumed that Fe tips have constant polarizations over the
range from -0.5eV to +0.5eV of $\sim$40\%. Shown in Fig. 3(f-j) and Fig.4(f-j) are the simulated SP-STM images
obtained for a tip with constant polarization $P_{t}=33$\%.

The simulations are for a surface region identical to the one in the experiment [Fig.3(a-e) and Fig.4(a-e)]. First
of all, we find that the overall corrugation is also largest at small bias magnitudes, whereas it is smaller at
larger bias magnitudes, in agreement with the experiments. Moreover, comparing the simulated SP-STM images and
averaged total line profiles with the experimental ones, very good agreement is seen at bias voltage magnitudes
smaller than $\sim$0.3V. For larger voltage magnitudes, however, simulated and experimental profiles do not agree:
it is clearly apparent that the simulated images show magnetic contrast at all bias voltages with contrast reversal
occurring at $V_S$ near -0.5V. In contrast, in the experimental data the magnetic contrast reversal occurs at +0.4
V.

Similar to experiment, we have also separated the magnetic and non-magnetic components from the total SP-STM
theoretical height profiles. Fig. 5(f-j) and Fig. 6(f-j) are the height profiles corresponding to the simulated
SP-STM images shown in Fig. 3(f-j) and Fig. 4(f-j), respectively.

Comparing non-magnetic profiles with the experimental ones, we find that both agree well showing a simple
sinusoid-like form with maximum amplitude at small bias magnitudes. The magnetic profiles also are in a good
agreement with experiment for biases between -0.4V and +0.2V, albeit the shape of the simulated profiles appears to
be more rounded. However, in contrast to experimental observations, simulated magnetic profiles do not exhibit
magnetic contrast reversal at positive biases, but rather show the reversal at negative bias -0.5V. Since these
simulated profiles correspond to an electronically featureless tip, i.e. depend only on the surface electronic
properties, we conclude that the experimental magnetic profiles cannot be adequately described solely in terms of
the surface electronic structure but that tip effects are essential and cannot be disregarded.

To check the sensitivity of the SP-STM profiles to the tip electronic properties, we have used the same theoretical
approach described above, but employed nonmagnetic and magnetic tip DOS $n^{t}\left(E\right)$ and
${m}^{t}\left(E\right)$ corresponding to the DOS of a single Fe atom on a Fe(110) surface extracted from
Ref.\cite{Heinze1}. The corresponding magnetic tip polarization $P_t$ as a function of energy for both tip models
is shown in Fig.7(a).

The experimental and theoretical magnetic and non-magnetic line profile magnitudes (peak to valley) are plotted vs.
$V_{s}$ in Fig. 7(b). Clearly, the magnitudes of the experimental non-magnetic contributions (Fig.7(b), curve 2)
are maximum at small voltage, whereas they get smaller at larger magnitudes of voltage. For the experimental
magnetic component (Fig.7(b), curve 1), the sign change can be clearly seen at +0.4V, indicating the change of
polarity of the magnetic contrast.

The theoretical magnetic and non-magnetic line profile magnitudes corresponding to the tip with constant DOS
(curves 3 and 4) and with Fe-atom tip (curves 5 and 6) are also shown in Fig. 7(b). For all biases and tip models,
the shape, amplitude, and periodicity of the non-magnetic profiles is well reproduced and has a simple
sinusoid-like form. At bias voltages from -0.2V to +0.2V the magnitude of these non-magnetic profiles is
essentially independent of the tip electronic properties; for larger biases the effect of the tip becomes more
pronounced, but the difference between different tips does not exceed $\sim$20\%. We conclude from this that the
specific tip electronic structure is not essential for the non-magnetic contribution to the SP-STM profiles.

The situation is remarkably different for magnetic profile data [Fig.7(b), curves 3 and 5]. As discussed before,
although the shape and periodicity of the magnetic profiles is reasonably well reproduced already with constant DOS
tip, the magnitudes and regions of magnetic contrast reversal differ substantially. Employing a more realistic tip
DOS severely modifies the magnitude of the magnetic component in simulated SP-STM profiles.

The magnetic profiles for Fe-tip are in remarkable qualitative agreement with experiment, since the contrast
reversal occurs at +0.4V, as with the experimental data. As can be clearly seen, however, the quantitative
agreement is far from being satisfactory, as the magnitude of the simulated magnetic profiles is significantly
underestimated for bias voltages larger than -0.6V. This underestimation is due to the specific magnetic structure
of the model Fe-tip[Fig.7(a)]. According to Eq.1, the tip DOS at positive energies affects the negative bias STM
current, while the tip DOS at negative energies affects the positive bias STM current. Therefore, the smoothly
growing Fe tip DOS at positive energies results in shift of the contrast reversal to more negative (less than
-0.8V) bias voltages and is not visible in the experimental data shown here; the underestimation of the magnitude
of the simulated magnetic profiles at positive biases occurs because the tip polarization at negative energies down
to -0.7V does not exceed 10\%. Although none of our tip models have provided excellent overall agreement with the
experimental magnetic data, we have clearly shown the importance of the specific tip electronic properties in the
SP-STM studies. To achieve better agrement between theory and experiment, one might need more precise information
regarding the experimental tip geometrical and electronic properties.

\subsection{Summary}
In summary, we have analyzed the bias-dependent magnetic profiles obtained by SP-STM on a model aFM surface at room
temperature. Particularly, we have shown that a polarization reversal occurs and that the magnetic contrast and
magnetic line profile shape vary with the bias voltage. Based on our spin-polarized DFT calculations, we have found
that the exact geometry and electronic structure of the tip atoms significantly affects the measured magnetic
profiles. For the nonmagnetic contribution, especially at low bias voltages, the effect of the tip electronic
properties is not essential, which might explain the success of constant tip DOS models in conventional
(non-magnetic) theoretical STM studies. In contrast, the magnetic part of the tunneling current is much more
sensitive to the specific tip electronic properties. Employing different types of tips can result in qualitatively
different results, which points to the importance of more accurate characterization of tip properties in magnetic
STM experiments.

\center{ACKNOWLEDGEMENTS}

This work has been supported by the National Science Foundation under grant Nos. 9983816 and 0304314.

\newpage
\begin{figure}
\caption {(a) is a top view of the bulk terminated surface model with magnetic moments indicated; (b) is the
geometrical configuration of the SP-STM experiment; The cross and dot represent the directions of the magnetic
moments.}
\end{figure}

\begin{figure}
\caption {SP-STM image of \etaMnN\ (010) surface acquired at V$_{s}$ = -0.2 V and I$_{t}$ = 0.3 nA using an
Fe-coated W tip; Up arrows correspond to the bright rows, down arrows correspond to the less bright rows; The
ellipse indicates one of the defect regions. The grey-scale is 0.28 \AA .  }
\end{figure}

\begin{figure}
\caption { Experimental and theoretical SP-STM images for negative bias voltages. (a-e) A series of SP-STM images
acquired using a Fe-coated W tip corresponding to the same box region in Fig.2, with corresponding height line
profiles. V$_{s}$ from -0.8 V to -0.1 V, and I$_{t}$=0.3 nA. (f-j) Corresponding theoretical images calculated
using spin-polarized DFT assuming tip with constant DOS. All the line profiles are in the same scale. The maximum
height profile magnitude is 0.28\AA . }
\end{figure}

\begin{figure}
\caption { Experimental and theoretical SP-STM images for positive bias voltages. (a-e) A series of SP-STM images
acquired using a Fe-coated W tip corresponding to the same box region in Fig.2, with corresponding height line
profiles. V$_{s}$ from +0.1 V to +0.8 V, and I$_{t}$=0.3 nA. (f-j) Corresponding theoretical images calculated
using spin-polarized DFT assuming tip with constant DOS. All the line profiles are in the same scale. The maximum
height profile magnitude is 0.28\AA .}
\end{figure}

\begin{figure}
\caption {Resulting magnetic and non-magnetic height profiles for negative bias voltages. (a-e) Experimental
non-magnetic (black) and magnetic (red) profiles. (f-j) Corresponding theoretical non-magnetic (purple) and
magnetic (blue) height profiles calculated using spin-polarized DFT assuming tip with constant DOS. }
\end{figure}

\begin{figure}
\caption {Resulting magnetic and non-magnetic height profiles for positive bias voltages. (a-e) Experimental
non-magnetic (black) and magnetic (red) profiles. (f-j) Corresponding theoretical non-magnetic (purple) and
magnetic (blue) height profiles calculated using spin-polarized DFT assuming tip with constant DOS. }
\end{figure}

\begin{figure}
\caption{  (a) Tip polarization as a function of sample bias voltage as used in the theoretical tip models. Data
for Fe tip is extracted from Ref.~\cite{Heinze1}. (b) The magnitude of magnetic component (red triangles left -
exp, green circles - Fe-tip model, black triangles right - constant DOS tip model) and non-magnetic component (red
squares - exp, green triangles up - Fe-tip model, black circles - constant DOS tip model). Fitted solid curves 1
and 2 (exp) and dashed curves 3-6 (theory) are shown to guide the eye.}
\end{figure}

\end{document}